\documentclass[a4paper,fleqn,usenatbib]{mnras}

\usepackage{txfonts}

\usepackage[T1]{fontenc}
\usepackage{ae,aecompl}

\usepackage{graphicx}	
\usepackage{siunitx}
\usepackage{amssymb}	
\usepackage{amsmath}	
\usepackage{color}

\usepackage{caption}
\usepackage{subcaption}
\graphicspath{{./figs/}}

\title[Collective vortex motion with a pinning barrier]{Collective, glitch-like vortex motion in a neutron star with an annular pinning barrier}

\author[J. R. L\"onnborn et al.]
{
J. R. L\"onnborn,$^{1}$\thanks{E-mail: lonnbornj@student.unimelb.edu.au}
A. Melatos$^{1}$\thanks{E-mail: amelatos@unimelb.edu.au}
and B. Haskell$^{2}$\thanks{E-mail: bhaskell@camk.edu.pl}
\\
$^{1}$School of Physics, University of Melbourne, Parkville, VIC 3010, Australia \\
$^{2}$Nicolaus Copernicus Astronomical Center of the Polish Academy of Sciences, Ulica Bartycka 18, 00-716 Warszawa, Poland
}

\date{Accepted XXX. Received YYY; in original form ZZZ}

\pubyear{2018}

\begin{document}
\label{firstpage}
\pagerange{\pageref{firstpage}--\pageref{lastpage}}
\maketitle

\begin{abstract}
Neutron star glitches are commonly believed to occur, when angular momentum is transferred suddenly from the star's interior to the crust by the collective unpinning and repinning of large numbers of superfluid vortices. In general, the pinning potential associated with nuclei in the crustal lattice varies as a function of radius. We explore vortex dynamics under these conditions by solving the three-dimensional Gross-Pitaevskii equation in a rotating, harmonic trap with an axisymmetric `moat' of deeper pinning sites on an otherwise uniform, corotating pinning grid. The moat is designed to resemble crudely a radially dependent pinning profile in a neutron star crust, although the values of the pinning potential are not astrophysically realistic due to computational constraints. It is shown that vortices accumulate in the moat, inducing large differential rotation which can trigger mass unpinning events. It is also shown that the system self-adjusts, such that the net vortex flux out of the system is the same with and without a moat, as the trap spins down, but glitches are less frequent and larger when the moat is present. The results, generated for an idealized system, represent a first step towards including stratified pinning in quantum mechanical models of neutron star glitches.
\end{abstract}




\section{Introduction}

The standard composition of the inner crust of a neutron star is a lattice of nuclei immersed in a sea of superfluid neutrons and degenerate electrons \citep{Baym1971}. The superfluid nucleates vortices as the star rotates, each carrying a quantum of circulation $\kappa = h/m$, where $m=2m_n$ is the mass of a Cooper pair. For the densities found in the inner crust [$4 \times 10^{11} \lesssim \rho / (\si{g.cm^{-3}}) \lesssim 1 \times 10^{14}$], first principles calculations suggest that vortices pin at or between nuclei in the lattice \citep{Avogadro2008,Chamel2008}. As the star spins down, vortex pinning prevents the superfluid from decelerating with the crust, generating a rotational shear. When the shear reaches a critical value, vortices unpin and transfer their angular momentum to the crust, causing a spasmodic increase in the star's rotational frequency known as a glitch \citep{Edlin1975}. Many (typically $10^7-10^{15}$) vortices unpin simultaneously, triggered by various collective knock-on mechanisms \citep{Warszawski2012a}.\par
In the absence of pinning, vortex-vortex repulsion (due to the Bernoulli force) is optimized in a triangular Abrikosov lattice \citep{Tkachenko1966}. The addition of pinning sites distorts this configuration, as vortices self-organize to balance competition between inter-vortex repulsion and attractive or repulsive pinning interactions. The equilibrium is frustrated in general. Frustrated systems have been studied in the context of terrestrial Bose-Einstein condensates (BECs) by superposing a corotating square optical lattice on a triangular vortex lattice \citep{Tung2006}. In the astrophysical context it has been shown that frustration due to vortex-flux-tube pinning in a neutron star's outer core leads to superfluid turbulence and microscopic vortex tangles \citep{Drummond2017a,Drummond2017}.\par
The radius and spacing of nuclei in the lattice and the sign of the vortex-lattice interaction depend on density, leading naturally to the suggestion that the strength of pinning varies between different regions of the crust \citep{Negele1973,Alpar1977,Alpar1984,Donati2004,Donati2006a}. Until recently, calculations of the vortex-lattice interaction have been semi-classical, based either on Ginzburg-Landau theory \citep{Epstein1988} or the Thomas-Fermi ansatz in the local density approximation \citep{Donati2004,Donati2006a}. Lately calculations have also been done using Hartree-Fock-Bogoliubov mean-field theory \citep{Avogadro2007,Avogadro2008}, focusing on mesoscopic interactions between a vortex and many pinning sites rather than computing the microscopic force per pinning site \citep{Seveso2016}.
\par
In this paper we study pinning in the situation, where a ring-like barrier (`moat') of deeper pinning sites at some fixed radius is superposed on a uniform lattice. The aim is to simulate, in an idealized fashion, the density-dependent, stratified pinning in a neutron star proposed by previous authors \citep{Alpar1977,Anderson1982,Epstein1988,Donati2004,Donati2006a}. A similar scenario is studied by  \cite{Sedrakian1999}, who consider vortex accumulation, collective vortex cluster interactions, and glitch generation in the presence of a potential barrier at the crust-core interface. The central -- and subtle -- physical question addressed by the present paper is: does the moat present a heightened barrier to outward vortex motion, as the star spins down? Or does the vortex array self-adjust to nullify the moat, i.e. do vortices pin preferentially in the moat, increasing the Magnus force locally and thereby lowering the barrier? If self-adjustment occurs, is it complete, or does the moat leave an imprint on vortex motion and glitch statistics?
\par 
The paper is organized as follows. We build an idealized Gross-Pitaevskii model of a decelerating, pinned BEC and study outward vortex drift and vortex avalanche dynamics with and without a moat. In Section \ref{model} we describe the Gross-Pitaevskii model and its limitations, specifically its idealized form and astrophysically unrealistic parameter choices (imposed by computational constraints). In Section \ref{sec:equilibrium} we compute the density and velocity fields for representative configurations, with and without a moat, in equilibrium. Section \ref{flux} compares the outward vortex flux for moats of various depths, as the trap spins down. It is shown that large differential rotation can develop in the vicinity of the moat, and that the vortex array self-adjusts such that the outward vortex flux is approximately unchanged compared to when the moat is absent. In Section \ref{dynamics} we present evidence of glitches in the simulations and calculate their size and waiting-time statistics, generalizing previous studies without a moat.
\section{Gross-Pitaevskii simulations} \label{model}
\subsection{Stratified pinning} \label{prev-work}
The configuration of superfluid vortices in a nuclear lattice depends on the pinning energy $E_{p}$, the energy difference between the non-interacting configuration (where the vortex-nucleus separation is large) and the zero-distance configuration (where the vortex core coincides with a nucleus). Positive $E_p$ means that vortex-nucleus pinning is energetically favourable, while negative $E_{p}$ favors interstitial pinning, which maximizes vortex-nucleus separation. A third possibility occurs when a vortex core is larger than a Wigner-Seitz cell in the nuclear lattice, so that the distinction between nuclear and interstitial pinning breaks down \citep{Donati2006a}.\par
Table \ref{tab:values} presents values of $E_{p}$ for different densities calculated by various investigators under conditions relevant to a neutron star. Various physical inputs and calculational schemes have been employed, including a homogeneous `liquid drop' model, where the difference in condensation energies is considered \citep{Alpar1984}; a phenomenological approach based on Ginzburg-Landau theory \citep{Epstein1988}; a semiclassical model based on the Thomas-Fermi ansatz in the local density approximation \citep{Donati2006a}; and the fully quantum Hartree-Fock-Bogoliubov mean field theory in the Wigner-Seitz approximation \citep{Avogadro2008}. \cite{Chamel2007} compared the Wigner-Seitz approximation to a full band theoretic model of dense neutron star matter. They found that the Wigner-Seitz approximation is well suited to the higher temperatures of young neutron stars and during core-collapse supernovae but it breaks down at lower temperatures ($T \lesssim 0.1\, \si{MeV}$), where entrainment becomes important. In all cases the crustal composition comes from the work of \cite{Negele1973}.
\par
\begin{table*}
	\centering
	\caption{Pinning energy $E_{p}$ calculated by various authors for different mass densities $\rho$. Note that $E_{p}$ is not a monotonic function of density: the end points of the density range do not correspond to the extrema of $E_p$. Sign convention: $E_{p}>0$ for nuclear pinning, $E_{p}<0$ for interstitial pinning.}
	\label{tab:values}
	\begin{tabular}{lllll} 
		\hline	
		Authors & $\rho \,(10^{13}\,\si{g.cm}^{-3})$ & $E_{p} \,(\si{MeV})$ & Physics & Calculational method \\
		\hline
		\cite{Alpar1984} 	& $[3,13]$ & $[0.5,3]$ & Liquid drop & Difference in condensation energies\\
		\cite{Epstein1988} 	& $[0.07,12.6]$ & $[-2.5,15]$ &	Phenomenological & Ginzburg-Landau\\
		\cite{Donati2006a} 	& $[0.15,13]$ & $[-0.81,3.38]$ & Semiclassical	& Thomas-Fermi \\
		\cite{Avogadro2008} & $[0.17,6.2]$ & $[-6.21,5.03]$ & Full quantum	& Hartree-Fock-Bogoliubov with \\
						& $[0.17,6.36]$ & $[-18.27, 3.85]$ & & Wigner-Seitz approximation\\
		\hline
	\end{tabular}
\end{table*}
These results are extended by calculations of the pinning potential per unit vortex length, which take into account the rigidity of the vortex and the fact that it interacts with a lattice \citep{Seveso2016,Wlazowski2016a}. \cite{Seveso2016} found weaker pinning compared to calculations involving a single pinning site but concluded that the largest pinning forces are still sufficient to store enough angular momentum in the crust to explain large glitches, such as those observed in the Vela pulsar. \cite{Wlazowski2016a} solved the time-dependent Hartree-Fock-Bogoliubov equations and showed that the pinning force is repulsive (resulting in interstitial pinning) and that its magnitude increases with density in the range $1.4 \times 10^{13} < \rho / (\si{g.cm^{-3}}) < 3.1 \times 10^{13}$.
\par
The pinning strength is in general a non-monotonic function of density and hence of radius. \cite{Alpar1984} found (Table \ref{tab:values}) that $E_p$ grows with increasing density to a maximum of $3 \, \si{MeV}$ at $\approx 7 \times 10^{13}\,\si{g.cm^{-3}}$ and falls off at the base of the inner crust. \cite{Epstein1988} reported a larger $E_\text{p}$ range, with the extrema ($-8.2 \,\si{MeV}$ and $15 \, \si{MeV}$) both lying towards the middle of the density range, and weaker pinning at the top and bottom of the inner crust. \cite{Donati2006a} found significant nuclear pinning ($2.5 < E_\text{p}/\si{MeV} < 3.5$) only in the layer $2 \times 10^{13} < \rho / (\si{g.cm^{-3}}) < 5 \times 10^{13}$, with negligible pinning elsewhere. The presence of $E_\text{p}$ extrema in the inner crust motivates the study in this paper, where local pinning takes a higher value in an annular moat.
\subsection{Numerical model} \label{model-detail}
Following previous work \citep{Warszawski2011a,Warszawski2012a,Melatos2015,Drummond2017a,Drummond2017}, we study vortex pinning in a neutron star computationally by modelling the system as a weakly interacting BEC in a rotating, decelerating, harmonic trap and solving the time-dependent Gross-Pitaevskii equation (GPE) on a three-dimensional grid [see \cite{Simula2008} and \cite{Schneider2006} for numerical details]. There are many reasons why this model is highly idealized. For example, the ratio of pinning sites to vortices in our simulations is of order $10$, rather than $10^{10}$ in a neutron star; the linear dimensions of the simulation box are $\sim 1\,\si{fm}$ (see Table \ref{tab:params}); the neutron superfluid in a neutron star is a strongly interacting fermionic condensate rather than a dilute, weakly interacting Bose gas, and so on. These limitations are discussed thoroughly by \cite{Haskell2015b} and in Section $7$ in \cite{Drummond2017}. However, the model is computationally tractable and has a successful record of capturing the collective knock-on processes which cause the scale-invariant behaviour of superfluid vortex avalanches under neutron star conditions \citep{Warszawski2011a,Warszawski2012a,Warszawski2012,Warszawski2013}. \par
In the frame corotating with the trap at angular velocity $\Omega$, the condensate order parameter $\psi(\mathbf{r},t)$ is described by the dimensionless stochastic GPE \citep{Gardiner2002a},
\begin{equation}\label{GPE}
(i-\gamma)\frac{\partial \psi}{\partial t} = -\frac{1}{2} \nabla^2 \psi + \big( V + \lvert \psi \rvert ^2 \big)\psi - \Omega \hat{L}_z\psi + i \gamma \mu \psi.
\end{equation}
Here $\mu$ is the chemical potential, and the term $-\gamma \partial \psi / \partial t$ ($\gamma \propto T$) models dissipation of sound waves by a viscous thermal cloud; see \cite{Warszawski2011a} for details. Decreasing $\gamma$ increases the decay time-scale of acoustic pulses emitted by moving vortices, which can unpin further vortices and trigger avalanches \citep{Warszawski2012a}. In this paper we take $\gamma = 0.1$, except where noted in Section \ref{sec:stats}. Note that $\psi$ is normalized such that the total number of bosons, $N_0$, equals $\int d^3 \mathbf{r} \, \lvert \psi(\mathbf{r},t) \rvert^2$. Length, time and energy in (\ref{GPE}) are given in units of $\hbar/(m\tilde{n}_0g)^{1/2}$, $\hbar / (\tilde{n}_0g)$ and $\tilde{n}_0g$ respectively, where $g$ is the boson coupling constant and $\tilde{n}_0$ is the mean boson density.\par
The angular velocity of the trap is updated self-consistently from one time-step to the next according to
\begin{equation}
I_\mathrm{c} \frac{d\Omega}{dt} = - \frac{d \langle \hat{L}_z\rangle}{dt} + N_\mathrm{ext},
\end{equation}
where $I_\mathrm{c}$ is the moment of inertia of the crust, $N_\mathrm{ext}$ is the braking torque (of electromagnetic origin in a neutron star), and $\langle \hat{L}_z\rangle = \langle \psi \lvert \hat{L}_z \rvert \psi \rangle$ is the expectation value of the angular momentum of the condensate, which responds to changes in vortex positions. In this paper we take $N_\mathrm{ext} = -0.005$ (dimensionless), except where noted in Section \ref{sec:stats}.\par
\begin{figure}
	\includegraphics[width=\columnwidth]{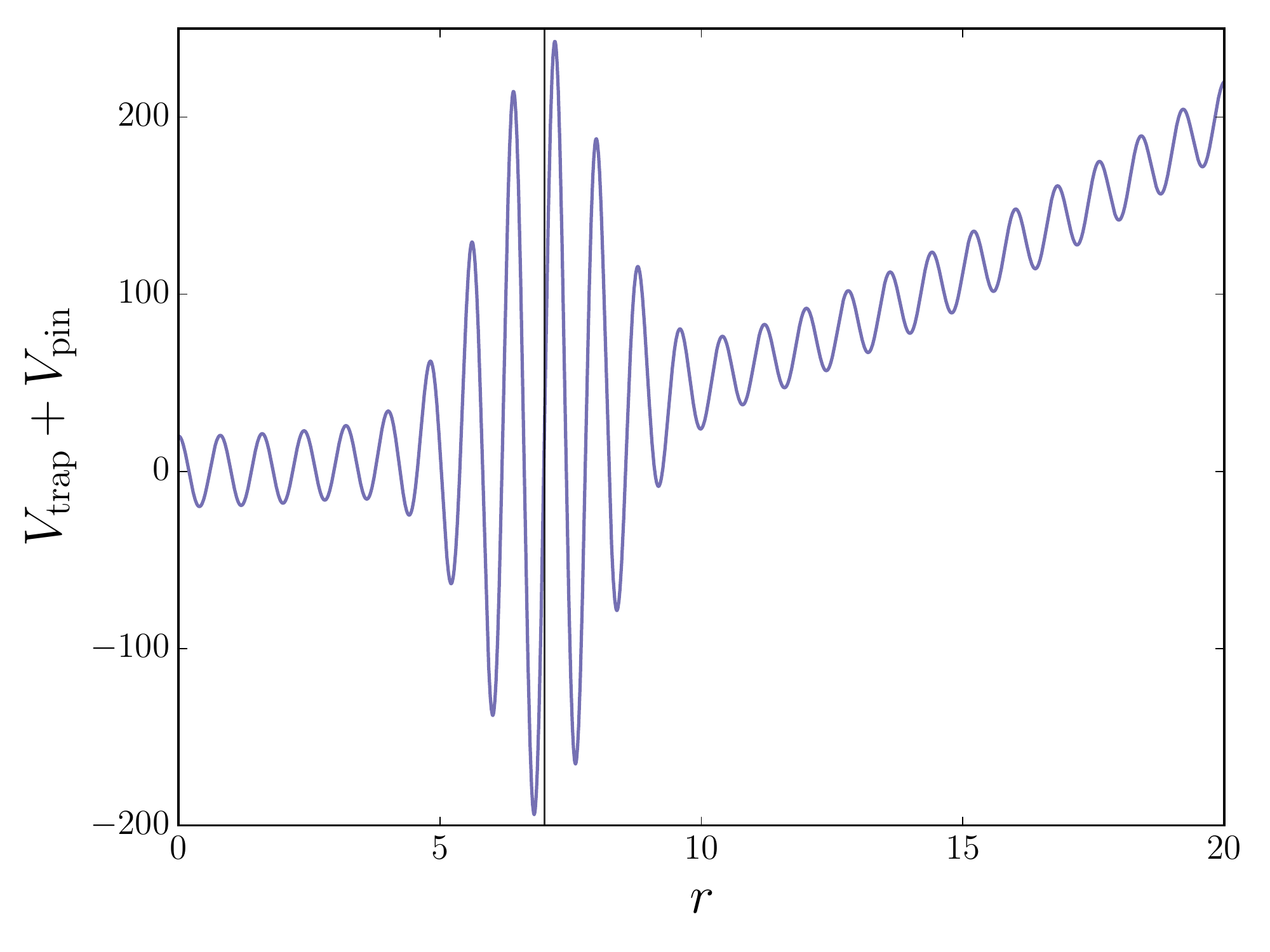}
    \caption{Harmonic trap and pinning potential as a function of radius $r$ with $V_\text{trap} = 0.5(x^2+y^2+\beta z^2)$, $\beta = 32$ (`pancake' geometry), $V_0 = 200$, $V_1 = 20$, $\xi = 1.6$, and $a_0 = 0.4$. The moat is centred on the vertical gray line at $R=7$.}
    \label{fig:potential}
\end{figure}

In the corotating frame there is a regular periodic lattice of pinning sites representing the crust. The pinning potential $V_\text{pin}$ increases to a maximum at a fixed radius $r=R$:
\begin{equation}\label{pin}
V_\text{pin} = \Big\{V_1 + V_0 \exp \Big[ - \frac{1}{\xi^2} \big(\sqrt{x^2+y^2}-R \big)^2 \Big]\Big\}\cos\bigg(\frac{\pi x}{a_0}\bigg)\cos\bigg(\frac{\pi y}{a_0}\bigg).
\end{equation}
In (\ref{pin}), $V_0$ and $V_1$ (both positive) are constants setting the strength of the pinning, and $a_0$ and $\xi$ set the lattice separation and width of the moat respectively. In (\ref{GPE}), the potential is $V=V_\text{trap}+V_\text{pin}$, where $V_\text{trap}$ is the harmonic trapping potential which confines the condensate, and the depth of the moat is controlled by the ratio $V_0/V_1$. Figure \ref{fig:potential} shows a representative example of $V$ versus $r$; $V_\text{trap}$ is specified in the caption.
\subsection{Neutron star parameters} \label{parameters}
In this section, we discuss critically the numerical values chosen for the parameters of the model. As noted above, computational limitations make it impractical to perform simulations under realistic neutron star conditions. Table \ref{tab:params} compares the values of parameters adopted in a typical simulation to those in a neutron star. The dimensional simulation parameters quoted in the table have been calculated by scaling their dimensionless counterparts by the units given after (\ref{GPE}). The choices are justified qualitatively below. \par
In Table \ref{tab:params}, $t_\mathrm{total}$ is the total spin down time, $t_\mathrm{cross}$ is the sound-crossing time, and $d_\mathrm{vort}$ is the intervortex separation. Of these, $t_\mathrm{total}$ is fixed by $\Omega$ and $N_\mathrm{ext}/I_\mathrm{c}$, which we discuss below, while $t_\mathrm{cross}$ is given by $\sim 2R(m/\tilde{n}_0g)^{1/2}$. The sound speed in a neutron star is approximately $c/\sqrt 3$, so that $t_\mathrm{cross}$ is many times shorter than $t_\mathrm{total}$. This is also true for the simulations. The pinning site separation $a_0$ in a simulation is small compared to $d_\mathrm{vort}$, as in a neutron star, but must be large enough for changes in the potential across a pinning site to be computationally resolvable. $V$ is chosen large enough to perturb the equilibrium vortex configuration but small enough for knock-on processes to trigger avalanches.
\par
The damping strength is given by $\gamma = 4mga^2k_BT/\pi\hbar$ in a dilute-gas Bose-Einstein condensate, where $a$ is the scattering length, $m$ is the mass of a Cooper pair, $k_B$ is Boltzmann's constant, $T$ is the condensate temperature and $g\approx 3$ is a correction factor \citep{Gardiner2002a}. The effective value of $\gamma$ in a neutron star, where the condensate is fermionic and strongly interacting, is unclear. Empirically speaking, though, if the glitches observed in pulsars are caused by vortex avalanches, then $\gamma$ should be small enough for avalanches to propagate, i.e. $\gamma \lesssim 0.2$. In practice, $\gamma$ should also be large enough that sound waves are damped out fast enough for the condensate to remain numerically stable, i.e. $\gamma\gtrsim 0.01$.
\par
The absolute values of $N_\mathrm{ext}$, $I_\mathrm{c}$ and $\Omega$ are irrelevant in determining the overall behaviour of the system. What matters is the characteristic time-scale over which the trap spins down, viz. $I_\mathrm{c}\Omega/N_\mathrm{ext}$. On the one hand, the spin-down time-scale should be long compared to the sound-crossing time of the system, so that the vortex array evolves through a sequence of metastable pinned states. This is true for both columns in Table \ref{tab:params}. It should also be long compared to the mean waiting time between avalanches, so that each avalanche amounts to a small fraction of the spin down, and the system loses vortices in a trickle (like in a neutron star) not in a rush. We wish the system to resemble, as closely as possible and for a large portion of the simulation time, a many-vortex system, to gain as much insight as possible into the collective multi-vortex unpinning physics. This is satisfied by the simulations studied in this work, which typically contain $100$--$200$ vortices initially. On the other hand, practically speaking, $N_\mathrm{ext}$ should be large enough so that multiple avalanches are triggered in a computationally reasonable time.
\begin{table}
	\centering
	\caption{Model parameters in dimensional form, in a typical simulation and in a neutron star. The parameter values in a neutron star are uncertain; the purpose of the table is to illustrate the relative scales.}
	\label{tab:params}
	\begin{tabular}{lll}
		\hline
	 	Quantity	 & 	Simulation 	 		&		Neutron star \\
		\hline
		$R$			 &	 	$10^{-15}\, \si{m}$		&	 $10^4\,\si{m}$ 			\\
		$V$ 		 &   	$10^{4} \, \si{MeV}$  		 &        $10^0$--$10^1\,\si{MeV}$             \\
		$N_\text{ext}/I_\text{c}$    &  $10^{46}\,\si{rad.s^{-2}}$ &   $10^{-15}$--$10^{-10}\,\si{rad.s^{-2}}$ 		 \\
		$\Omega$     &      $10^{24}$\,\si{rad.s^{-1}}       & 	  $10^{-1}$--$10^2\,\si{rad.s^{-1}}$		        \\
		$t_\mathrm{total}$		&		$10^{-21}\,\si{s}$				&		$	10^{9}\,\si{s}	$		\\
		$t_\mathrm{cross}$	&		$5 \times 10^{-24}\,\si{s}$		&		$5\times10^{-5}\,\si{s}$	\\
		$d_\mathrm{vort}$		 &		 $10^{-16}\,\si{m}$ 			& 		$10^{-5}\,\si{m}$		\\
		$a_0$					& 		$5\times10^{-17}\,\si{m}$				&	  $10^{-14}$--$10^{-13}\,\si{m}$ \\  
		\hline
	\end{tabular}
\end{table}
\section{Representative equilibrium with a moat}\label{sec:equilibrium}
\subsection{Circulation and vortex pattern}
We now test how the moat modifies the equilibrium configuration of the vortex array. The system is driven firstly to its ground state by propagating in imaginary time ($t \rightarrow it$) with zero spin down. We then propagate the ground-state wavefunction in real time with non-zero spin down, and examine the configuration at a relatively early time, $t=5.0$.\par
\begin{figure}
\centering{
	\includegraphics[width=\columnwidth]{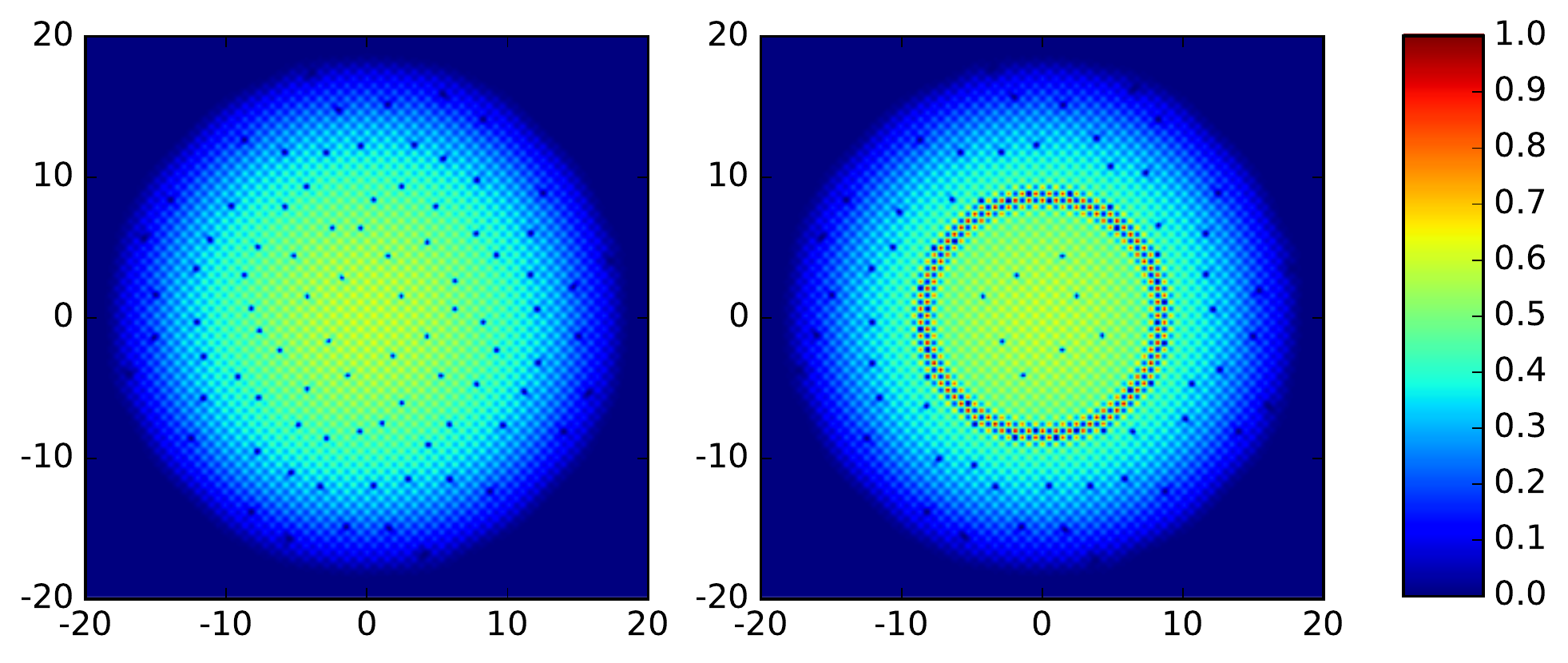}}
    \caption{Condensate density $\lvert \psi \rvert^2(x,y)$ (in arbitrary units) without (left panel) and with (right panel) a moat at $t=5.0$. Blue (red) represents low (high) density. Dark blue spots are vortices. Parameters: same as Figure \ref{fig:potential}.}
    \label{fig:density}

\end{figure}
\begin{figure}
	\includegraphics[width=\columnwidth]{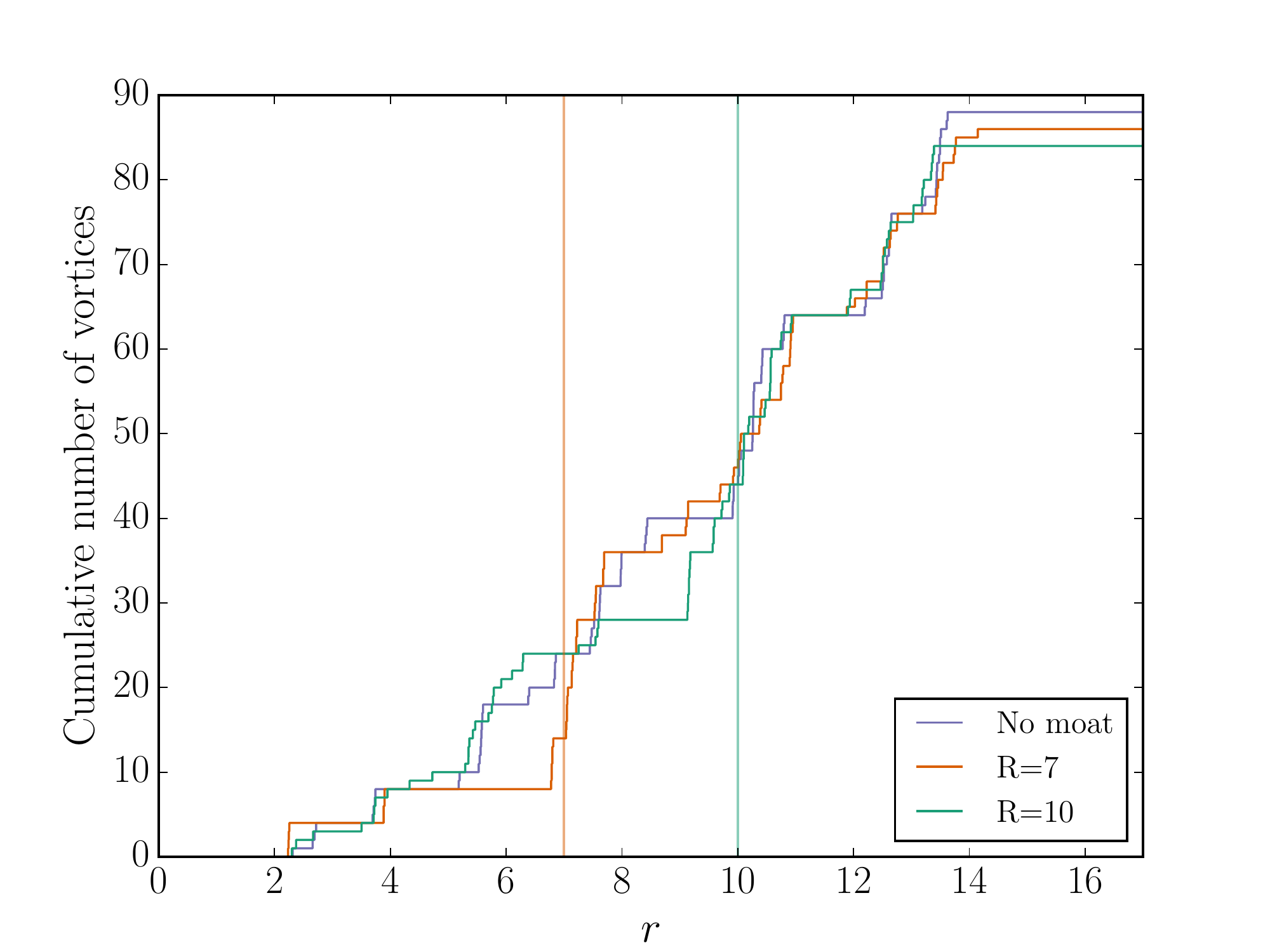}
    \caption{The cumulative number of enclosed vortices versus radial distance $r$ from the rotation axis for three different experiments: no moat (purple, $V_0=0$), moat at $R=7$ (orange), and moat at $R=10$ (green). The centres of the moats are indicated by the vertical orange and green lines. Vortices pin preferentially in the vicinity of a moat. Parameters: $t=1.0$, $V_0=200$, $V_1=20$, $\xi = 0.63$, $a_0=0.4$.}
    \label{fig:cumul_vort}
\end{figure}

Figure \ref{fig:density} plots the condensate density $\lvert \psi \rvert^2$ without (left panel) and with (right panel) a moat at $R=7$ with $V_0/V_1 = 10$. Dark blue spots in the condensate are vortices. In the left panel the density decreases away from the axis. In the right panel the condensate `pools' in the moat: the maximum of $\lvert \psi \rvert^2$ lies in the region $\lvert r-R \rvert < \xi$. This is a result of the pinning potential; the same effect is present with zero rotation and spin down. Figure \ref{fig:cumul_vort} plots the cumulative number of vortices enclosed within a radius $r=\sqrt{x^2+y^2}$, when there is no moat, and when the moat is centred at $R=7$ and $R=10$. Let the radial distance of the $i$-th vortex from the origin be denoted by $r_i$. For $R=7$, we have $2.23 \leq r_1,\,\ldots\,, r_8 \leq 3.90$, then a plateau in the graph until $6.78 \leq r_9,\,\ldots\,,r_{36} \leq 7.69$. Note that $r_{36}-r_9 = 0.91$ is of order the moat width $\approx \xi = 0.63$. The remaining vortices, which have $r_i > R + \xi$, lie in the range $11.90 \leq r_{37},\,\ldots\,, r_{86} \leq 14.15$. For $R=10$, we have $2.30 \leq r_1,\,\ldots\,, r_{28} \leq 7.59$ and $9.13 \leq r_{29}, \,\ldots\,, r_{64} \leq10.93$, so that approximately $42\%$ of vortices are pinned within $\lvert r-R \rvert < 2$ of the centre of the moat. The remaining vortices lie in the range $11.90 \leq r_{65},\,\ldots\,, r_{84} \leq 13.39$. In both cases, a large number of vortices are pinned in the vicinity of the moat.\par
\begin{table}
	\centering
	\caption{Number of vortices $N_\text{vort}$ inside ($\lvert r-R \rvert < \xi$) and outside ($\lvert r-R \rvert > \xi$) moats of different widths. The vortex overdensity (final column) is defined as $[n_\text{vort}(\lvert r-R \rvert< \xi)-\bar{n}_\text{vort}]/\bar{n}_\text{vort}$, where $n_\text{vort}$ is the local vortex density (number of vortices per unit area) and $\bar{n}_\text{vort}$ is $n_\text{vort}$ spatially averaged over the whole condensate. Parameters: $t=1.0$, $V_0/V_1 = 10$, $a_0=0.4$.}
	\label{tab:Nvort}
	\begin{tabular}{cccc}
		\hline
	 	$\xi/a_0$ & $N_\text{vort}(\lvert r-R \rvert< \xi)$ & $N_\text{vort}(\lvert r-R \rvert > \xi)$ & Overdensity\\
		\hline
		$1.58$ 	& $24$		& 	$61$	&	$2.16$	\\
		$2$ 	& $16$ 		&	$62$ 	&	$0.85$	\\
		$3$ 	& $16$	 	& 	$63$	&	$0.21$	\\
		$4$ 	& $32$ 		& 	$47$	&	$0.78$	\\
		\hline
	\end{tabular}
\end{table}
\par
Table \ref{tab:Nvort} shows the number of vortices pinned in moats of the same depth but different widths. Wider moats do not necessarily pin more vortices than narrower moats of the same depth. However, in all cases the vortex overdensity in the moat, which we define as $[n_\text{vort}(\lvert r-R \rvert< \xi)-\bar{n}_\text{vort}]/\bar{n}_\text{vort}$ where $n_\text{vort}$ is the local vortex density and $\bar{n}_\text{vort}$ is the spatially averaged $n_\text{vort}$, is greater than zero, indicating a greater concentration of vortices in the moat relative to the mean. \par
\begin{figure}
\captionsetup[subfigure]{aboveskip=-2pt}
\centering{
	\begin{subfigure}{0.4\textwidth}
	\includegraphics[width=\columnwidth]{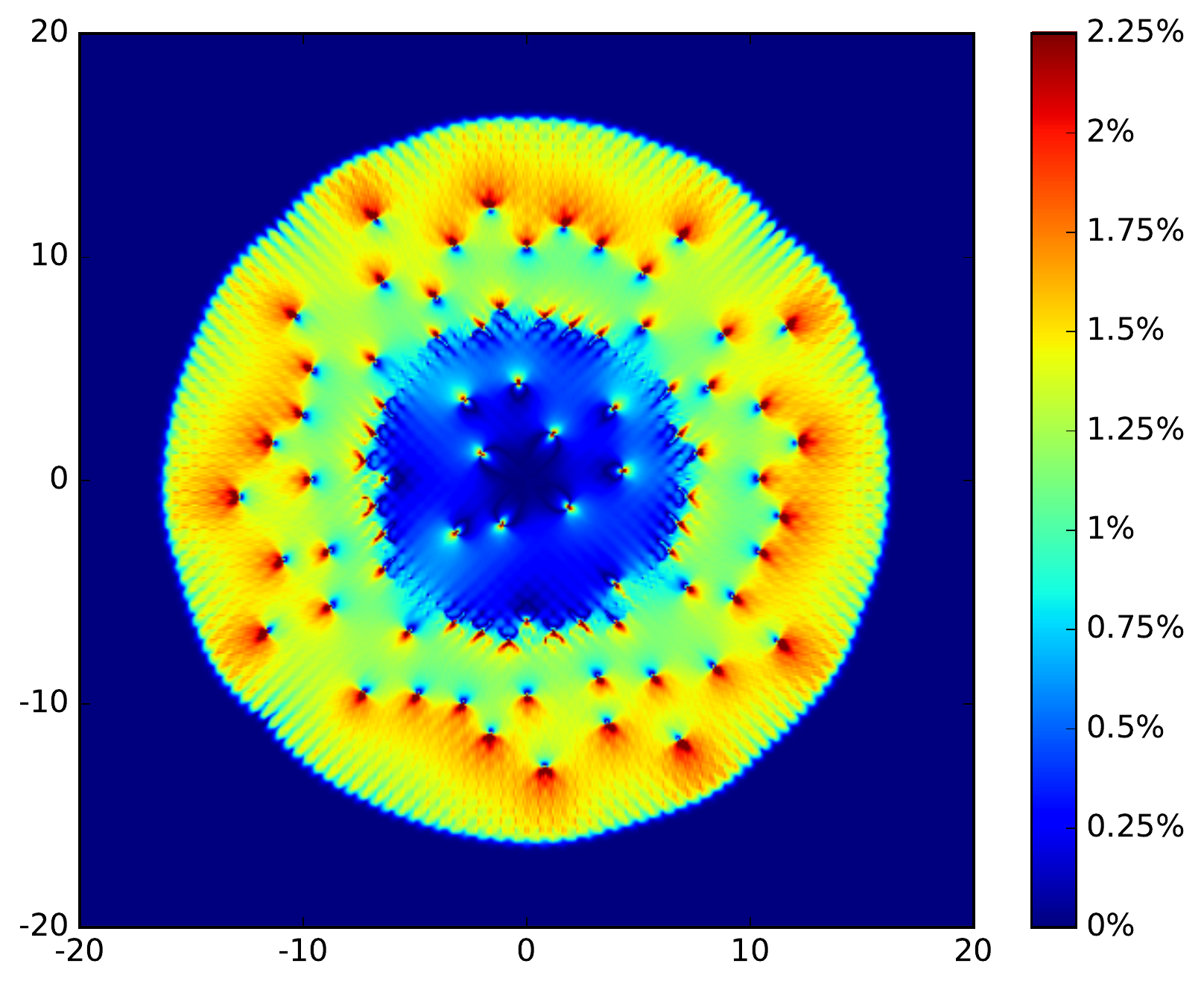}
	\caption{}
	\label{fig:flow_vars_a}
	\end{subfigure}
	\begin{subfigure}{0.45\textwidth}
	\includegraphics[width=\columnwidth]{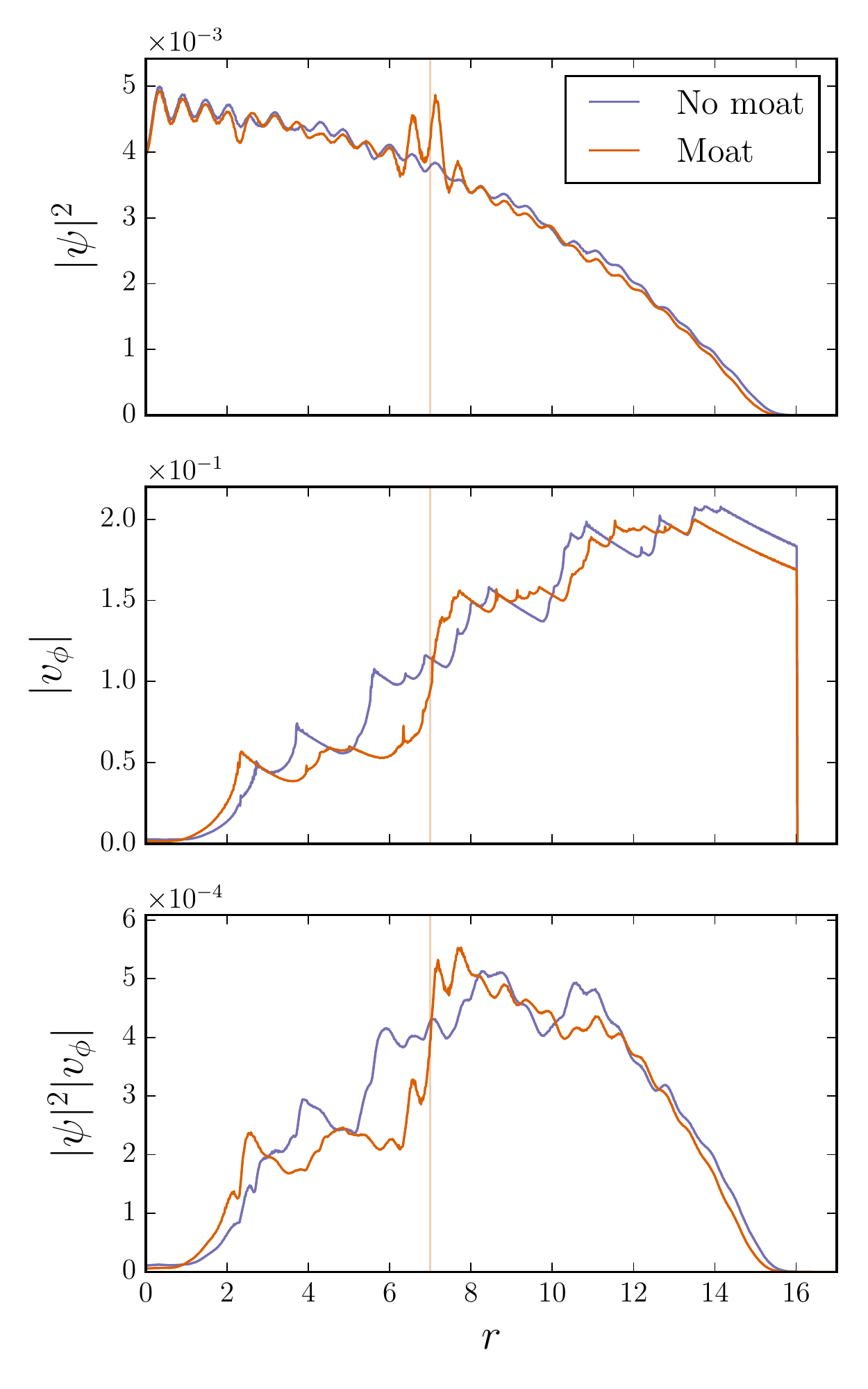}
	\caption{}
	\label{fig:flow_vars_b}
	\end{subfigure}
	\caption{Flow variables versus $r$ with and without a moat at $R=7$. (a) Contour plot of the magnitude of the azimuthal condensate velocity $\lvert v_{\phi} \rvert$ with a moat. The range of values represented in the contour plot is $0$ to $2.25\%$ of $\lvert v_\phi \rvert_{\text{max}}$; the numbers on the colorbar indicate a percentage of $\lvert v_\phi \rvert_{\text{max}}$. (b) Top to bottom: number density $n_0 = \lvert \psi \rvert^2$, $\hat{\phi}$ component of velocity $\lvert v_{\phi} \rvert$ and $\hat{\phi}$ component of mass current $\lvert \psi \rvert^2\lvert v_{\phi} \rvert$, as functions of radial distance $r$ from the rotation axis and averaged over $\phi$ with (orange) and without (purple) a moat. Parameters: $t=2.5$, $V_0/V_1 = 10$, $\xi = 0.63$, $a_0 = 0.4$.}
	\label{fig:flow_vars}
}
\end{figure}

The above observations suggest, that the circulation of the fluid (which is proportional to the number of vortices enclosed within radius $r$) is low for $r<R-\xi$ and high for $r>R+\xi$. We test this by computing the local fluid velocity
\begin{equation}
\mathbf{v} = -i\hbar (\psi^* \nabla \psi - \psi \nabla \psi^*)/(2m\lvert \psi \rvert^2).
\end{equation}
Figure \ref{fig:flow_vars_a} shows a contour plot of the magnitude of the azimuthal velocity component, $\lvert v_{\phi} \rvert$. A large number of vortices (identified by red spots, where the fluid velocity is high) are pinned near the moat at $R = 7$. This causes a large change in the velocity field: the low (blue) values inside the moat increase rapidly for $r \gtrsim R$. The range of velocities represented in the figure is $0$ to $2.25\%$ of $\lvert v_\phi \rvert_{\text{max}}$: all pixels with $\lvert v_\phi \rvert \geq 0.0225\lvert v_\phi \rvert_{\text{max}}$ are assigned the same (dark red) color. This is necessary to maintain contrast between pixels which are far from a vortex core, because the velocity field diverges inversely with distance from a vortex.
\par
Figure \ref{fig:flow_vars_b} plots flow variables of the condensate versus $r$, averaged over $\phi$. The top panel plots the condensate number density $n_0 = \lvert \psi \rvert^2$, which has a local maximum in the vicinity of the moat, seen also in the right panel of Figure \ref{fig:density}. The middle and bottom panels show $\lvert v_{\phi} \rvert$ and the azimuthal current density $\lvert j_{\phi} \rvert = \lvert \psi \rvert^2 \lvert v_{\phi} \rvert$ respectively. As $\lvert \psi \rvert^2$ takes particularly high values in the moat, we plot both $\lvert j_{\phi} \rvert$ and $\lvert v_{\phi} \rvert$ in order to verify that the local increase in $\lvert j_{\phi} \rvert$ near the moat is not just due to $\lvert \psi \rvert^2$ being higher there. This is important, because the Magnus force, which triggers unpinning, is proportional to $\lvert v_{\phi} \rvert$ not $\lvert j_{\phi} \rvert$.
\par
Let the relative change in a flow variable $X$ in the vicinity of the moat be defined as 
\begin{equation}
\frac{\lvert \Delta X \rvert}{X} = \frac{\max[X(\lvert r-R\rvert < \xi)] - \min[X(\lvert r-R\rvert < \xi)]}{\bar{X}(\lvert r-R\rvert < \xi)},
\end{equation}
where the overbar indicates a spatial average over values of $X$ in the region $\lvert r-R\rvert < \xi$, and $\min(\,\cdots)$ [$\max(\,\cdots)$] indicates we select the minimum (maximum) value of $X$ in $\lvert r-R\rvert < \xi$. Referring again to Figure \ref{fig:flow_vars}, we find $\lvert\Delta n_0\rvert/n_0=0.36,\,0.11$ and $\lvert\Delta j_{\phi}\rvert/j_{\phi}=0.75,\,0.11$ with and without a moat respectively. Larger relative changes occur in $n_0$ and $j_{\phi}$ where there is a moat. Additionally, we find $\lvert\Delta v_{\phi}\rvert/v_{\phi}=0.87$ and $\,0.20$ with and without a moat. Hence the mass current due to the moat is higher not only because $\lvert \psi \rvert^2$ takes a higher value there; $\lvert v_{\phi} \rvert$ is also higher. 
\subsection{Moat depth}\label{depth}
\begin{figure}
	\includegraphics[width=\columnwidth]{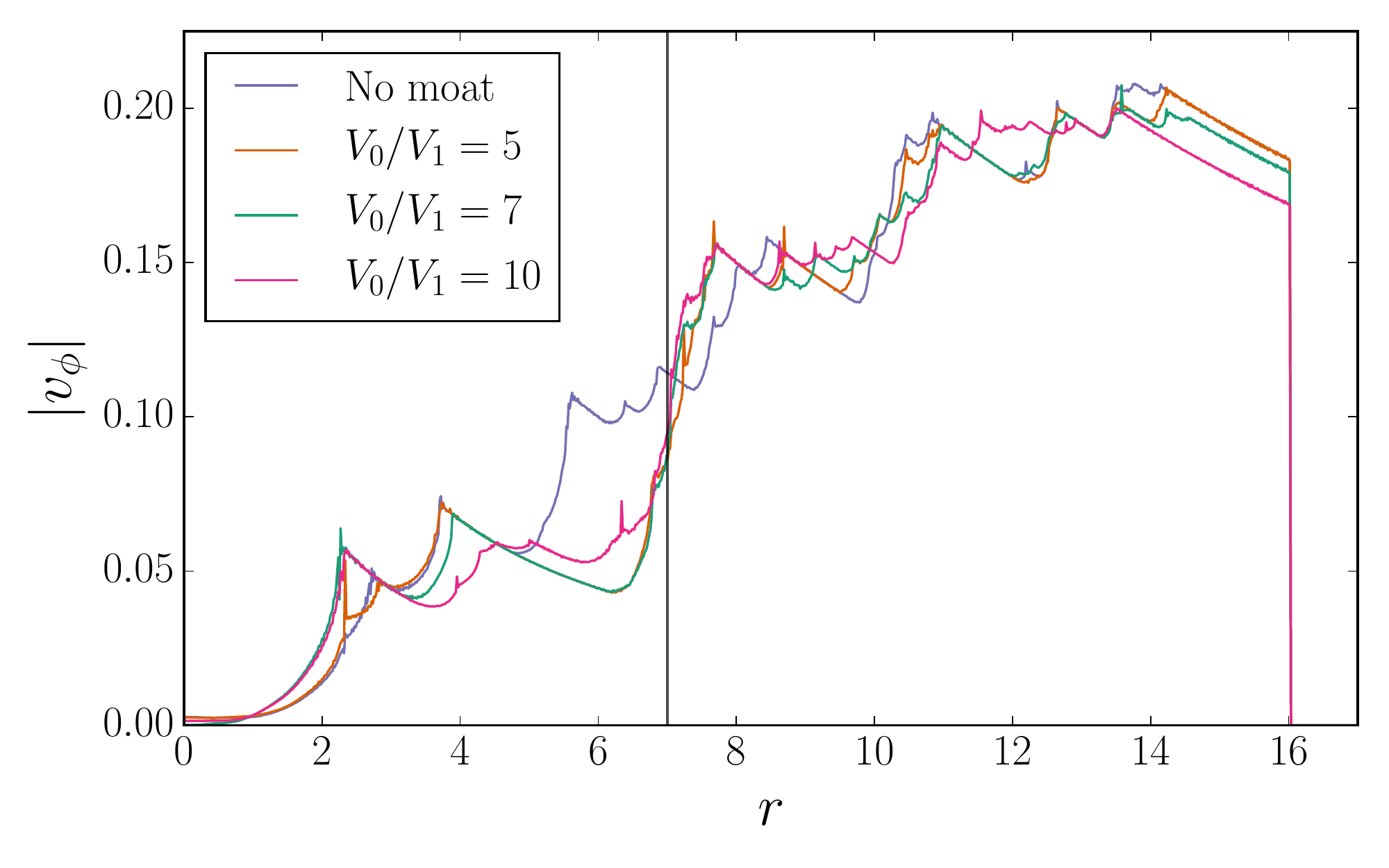}
    \caption{The azimuthal component of velocity $\lvert v_{\phi} \rvert$ (averaged over a circle) versus radius for moats of differing depth with $R=7$ (vertical gray line). $V_0/V_1$ quantifies the strength of pinning in the moat relative to the lattice, as defined in Section \ref{model-detail}. Parameters: same as Figure \ref{fig:flow_vars}.}
\label{fig:velocity_depth}
\end{figure}

In this section, we study how varying the depth of the moat, $V_0$, affects the vortex configuration in equilibrium. Figure \ref{fig:velocity_depth} shows $\lvert v_\phi(r) \rvert$ averaged over circles of constant radius for moats of various depths, with $R=7$ and $5 \leq V_0/V_1 \leq 10$. The figure shows an increase in $\lvert v_\phi \rvert$ across the moat, with $\lvert\Delta v_{\phi}\rvert / v_{\phi} = 1.13,\,1.09,\,0.87$ for $V_0/V_1 = 5,\, 7,\, 10$ respectively, compared to $\lvert\Delta v_{\phi}\rvert / v_{\phi} = 0.20$ for $V_0=0$. Every simulated moat produces a larger fractional increase in $v_\phi$ than no moat, but the fractional increase is larger for shallower moats with $5 \leq V_0/V_1 \leq 10$. \par
An alternative way to quantify the increase in $\lvert v_{\phi} \rvert$ across the moat is to smooth $\lvert v_{\phi} \rvert$ with a Savitsky-Golay (low-pass) filter of window size $0.85$ and take the gradient of the smoothed function at $r=R$. We find $d\lvert v_{\phi} \rvert/dr= 0.06,\, 0.15,\, 0.17$ for $V_0/V_1 = 5,\, 7,\, 10$ respectively, and $d\lvert v_{\phi} \rvert/dr = -0.003$ for $V_0=0$. According to this measure, deeper moats cause a steeper change in $v_{\phi}$. The results for both $\lvert\Delta v_{\phi}\rvert / v_{\phi}$ and $d\lvert v_{\phi} \rvert/dr$ are summarized in Table \ref{tab:depth}. The gradient indicates that the differential rotation and associated Magnus force are high. This is reminiscent of the `snowplow' model for giant Vela-like pulsar glitches, where a vortex sheet is initially pushed outwards, then released at the maximum of the density-dependent pinning force per unit length \citep{Pizzochero2011b}. In Section \ref{dynamics}, we explore the dynamical implications by searching for glitches in the simulation output and quantifying their statistics.\par
\begin{table}
	\centering
	\caption{Two measures of the effect of moat depth $V_0$ on the azimuthal velocity gradient in the moat: $\lvert\Delta v_{\phi}\rvert / v_{\phi}$ and $d \lvert v_{\phi} \rvert/dr$ (dimensionless) (defined in text). Parameters: same as Figure \ref{fig:flow_vars}.}
	\label{tab:depth}
	\begin{tabular}{lll} 
		\hline
		$V_0/V_1$ & $\lvert\Delta v_{\phi}\rvert / v_{\phi}$ & $d\lvert v_{\phi} \rvert/dr$ \\
		\hline
		$0$ 	& $0.21$ 	& $-0.003$ 	\\
		$5$		& $1.13$	& $0.06$ 	\\
		$7$		& $1.09$ 	& $0.15$ 	\\
		$10$ 	& $0.87$ 	& $0.17$	\\
		\hline
	\end{tabular}
\end{table}
The effect of a glitch is to correct accumulated stresses by transferring angular momentum to the crust. Hence a third way to quantify the effect of a moat is to compute the angular momentum. The total angular momentum of a system of vortices is given by
\citep{Fetter1965}
\begin{equation}\label{Lex1}
L = \pi \tilde{n}_0 \hbar \sum_i (Y^2-r_i^2),
\end{equation}
where $Y$ is the side length of the simulation box in the $x$ and $y$ directions and $i$ labels each vortex. Since the pinning potential affects the spatial distribution of vortices (Figure \ref{fig:cumul_vort}, Table \ref{tab:Nvort}) and hence $r_i$ in (\ref{Lex1}), we expect that a moat should alter $L$.
\par
We define the `excess' contribution to angular momentum from vortices outside the moat as
\begin{equation}\label{Lex2}
L_\text{ex} = L(r_i>R)-L_\text{NM}(r_i>R),
\end{equation}
where the subscript $\text{NM}$ denotes the no-moat system ($V_0=0$), and $r_i>R$ indicates that we include in $L$ only those vortices with $r_i>R$. The results for $R=5,\,10$ and $V_0/V_1 = 20,\, 10,\, 5,\, 2.5$ are shown in Table \ref{tab:L_excess}. We expect the numbers in Table \ref{tab:L_excess} to decrease down each column (deeper moats build up greater stresses) but remain positive (any moat builds up more stress than no moat). By and large these expectations hold except for two anomalous data points (both in the $R=5$ column). However, looking at a single time-step is insufficient here for the following reason. Suppose that we calculate $L_\text{ex}$ at some time $t=t_c$, and that in one simulation with large $V_0/V_1$ a glitch occured just prior to $t_c$, while in another simulation with small $V_0/V_1$ the most recent glitch was significantly earlier than $t_c$. A glitch corrects the build-up of $L_\text{ex}$. Hence $L_\text{ex}$ may be larger in the latter simulation despite the shallower moat; the two simulations are at different points in their cycle of building up and relaxing stress. We study glitches further in Sections \ref{flux} and \ref{dynamics}.
\begin{table}
	\centering
	\caption{Angular momentum induced by vortices outside the moat: $L_\text{ex}$ in units of $\pi \tilde{n}_0 \hbar$ [defined by Equations (\ref{Lex1}) and (\ref{Lex2})] for moats of various depths $V_0/V_1$. Parameters: $t=1.0$, $\xi =0.32$, $a_0=1$.}
	\label{tab:L_excess}
	\begin{tabular}{lll} 
		\hline
		$V_0/V_1$ & $L_\text{ex}(R=5)$ & $L_\text{ex}(R=10)$ \\
		\hline
		$20$ 	& $163.1$ & $326.6$ \\
		$10$	& $177.2$ & $179.3$ \\
		$5$		& $121.6$ & $119.5$ \\
		$5/2$ 	& $-21.7$ & $4.8$ \\
		\hline
	\end{tabular}
\end{table}
\section{Outward vortex flux during spin down} \label{flux}
This section looks at the effect of a moat on the spin down of the container. The motivation is partly astrophysical: we wish to know how a shell of stronger pinning affects the long-term deceleration of a neutron star's crust, even though it may be hard to disentangle from other spin-down effects in practice. We find in Section \ref{sec:equilibrium} that as vortices move radially outwards, they pin in the vicinity of the moat (Figure \ref{fig:cumul_vort}). In this respect, the moat acts like a divot or hole on a surface on which a sandpile is forming. Once some critical number of vortices pin near the moat (analogously, once the hole is filled with sand), the question becomes whether the outward vortex flux is the same as without a moat, or whether the flux is altered, retaining an imprint of the moat.
\par
\begin{figure}
	\includegraphics[width=\columnwidth]{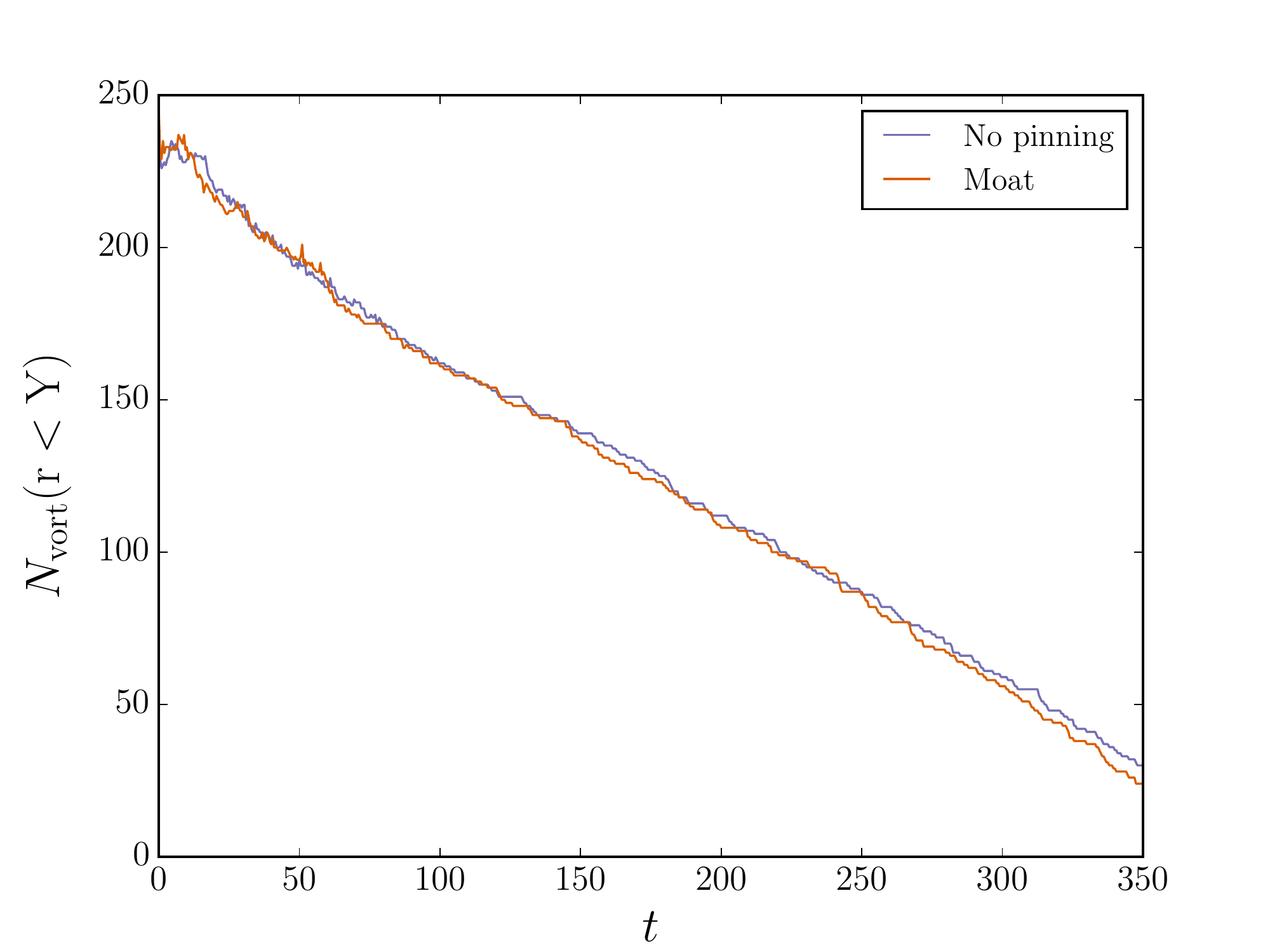}
    \caption{Total number of vortices versus time without pinning (purple curve; $V_0 = V_1 = 0$), and with a moat but no lattice (orange curve; $V_0 \neq 0$, $V_1=0$). Parameters: $V_\text{pin} = 0$ (purple curve); $V_0=200$, $V_1=0$, $R=10$, $\xi=1.58$, $a_0=1$ (orange curve).}
    \label{fig:nvort}
\end{figure}

Figure \ref{fig:nvort} graphs the total number of vortices in the system $N_\text{vort}(r<R)$ as a function of time for no pinning ($V_0=V_1=0$), and for a moat with pinning sites inside it but none outside it ($V_0 \neq 0$, $V_1=0$). Anticipating the study of glitch size and waiting time statistics in Section \ref{dynamics}, we investigate whether the total number of vortices and their distribution in the system, as functions of time, differ between the two cases. For $t\gtrsim 25$ vortices leave the system at an approximately constant rate in both simulations. There is little difference between the two curves: a linear fit to $N_\text{vort}(r<R)$ versus $t$ gives a gradient of $-0.55$ without pinning, and $-0.57$ with a moat, and the maximum difference between the number of vortices in each system is $11$ at $t=16$. If we think of the moat as a defect which perturbs the vortex distribution, then this result suggests that the vortex array self-adjusts to `heal' the defect: vortices pin near the moat, increasing the local Magnus force and lowering the barrier imposed by the moat, so that the outward vortex flux (after some period of equilibration) carries no imprint of the defect. \par
\begin{figure}
	\includegraphics[width=\columnwidth]{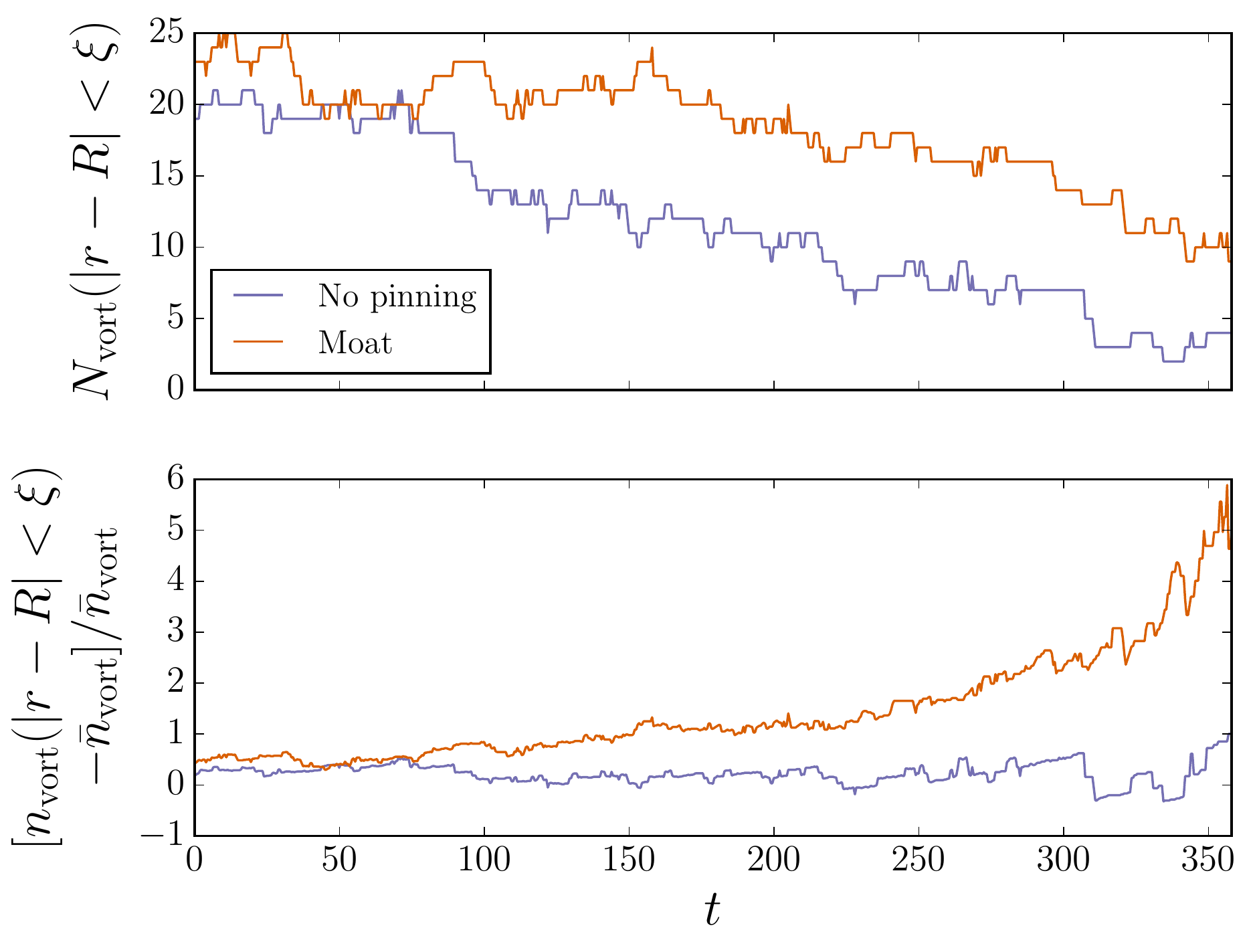}
    \caption{The spatial distribution of vortices versus time without pinning (purple curve; $V_0 = V_1 = 0$), and with a moat but no lattice (purple curve; $V_0 \neq 0$, $V_1=0$). Top panel: number of vortices in the region $\lvert r-R \rvert < \xi$. Bottom panel: overdensity of vortices in the region $\lvert r-R \rvert < \xi$, $[n_\text{vort}(| r-R |< \xi)-\bar{n}_\text{vort}]/\bar{n}_\text{vort}$, versus time. Parameters: same as Figure \ref{fig:nvort}.}
    \label{fig:nvort-2}
\end{figure}

\begin{figure}
\centering{
	\includegraphics[width=\columnwidth]{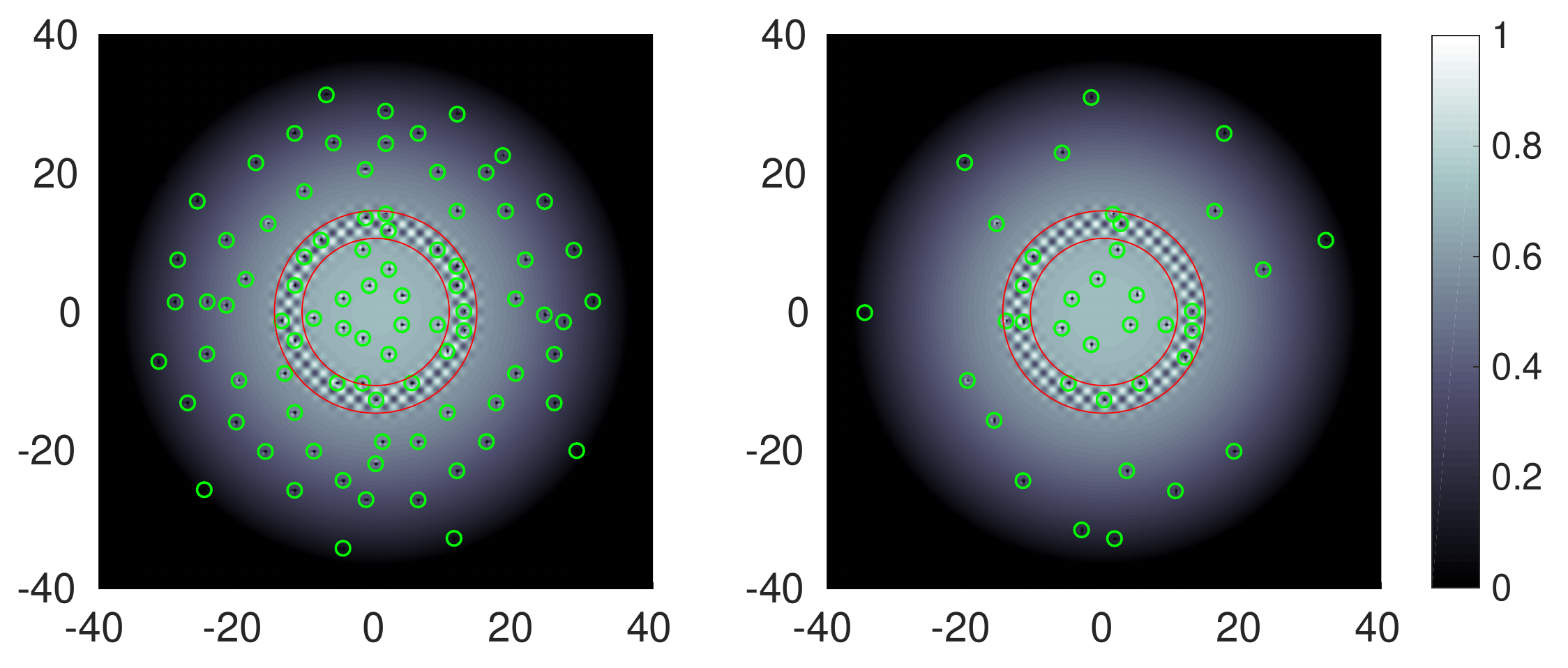}
    \caption{Alternative visualization of the vortex overdensity in the moat and its growth with time. Contour plots of the condensate density $\lvert \psi \rvert^2$ at (left) $t=250$ and (right) $t=331.45$. Light (dark) represents high (low) density, vortices are marked by open green circles, and the edges of the moat (at $R \pm \xi)$ are indicated by red circles. Parameters: same as Fgure \ref{fig:nvort}.}
    \label{fig:vort_density}
    }
\end{figure}

We now turn to the question of how the vortex pattern evolves in the presence of a moat, given that the net flux of vortices out of the system is unchanged from the no-moat configuration. Figure \ref{fig:nvort-2} shows the number of vortices pinned in the moat, $N_\text{vort}(\lvert r-R \rvert < \xi)$ (top panel), and the overdensity of vortices pinned in the moat, $[n_\text{vort}(| r-R |< \xi)-\bar{n}_\text{vort}]/\bar{n}_\text{vort}$ (bottom panel), as functions of time for the no pinning (purple) and moat (orange) configurations.
We see that the number of vortices pinned in the region $\lvert r-R \rvert< \xi$ is approximately constant until $t\approx 160$, when it begins to decrease linearly (orange curve, top panel). A linear fit to $N_\text{vort}(\lvert r-R \rvert< \xi)$ versus $t$ for $t>160$ gives a gradient of $-0.05$. However the density of vortices pinned in the moat relative to the system as a whole increases with time (orange curve, bottom panel). As the star spins down, both the net flux of vortices out of the moat and the system as a whole is positive, but vortices leave the system as a whole $\approx 10$ times faster than they leave the moat. This means that $N_\text{vort}(\lvert r-R \rvert< \xi)$ and $N_\text{vort}(r<Y)$ both decrease, the latter faster than the former. Figure \ref{fig:vort_density} shows a snapshot of $\lvert \psi \rvert^2$ at two different times, with vortices marked by open green circles. At $t=250$ (left) the vortex overdensity in the moat is $1.60$; by $t=331.45$ (right) it rises to $2.94$.
\section{Vortex avalanches}\label{dynamics}
We now study vortex avalanches and spasmodic spin down in order to investigate qualitatively how a moat affects neutron star spin-down and glitches. In Section \ref{sec:glitch-algorithm} we describe the algorithm used to find glitches in the spin-down data, and in Section \ref{sec:stats} we present glitch size and waiting time statistics.
\subsection{Glitch detection}\label{sec:glitch-algorithm}
In the glitch detection algorithm described by \cite{Warszawski2011a}, $\Omega(t)$ is first smoothed with a top-hat window function of width $t_\mathrm{sm}$ to combat numerical jitter. A glitch is deemed to occur at time-step $i$, whenever we have $\Omega(t_{i+1}) > \Omega(t_i)$, that is, whenever the smoothed angular velocity of the condensate increases. Let the end of a glitch $t_f$ be the first time-step after $t_i$ for which $\Omega(t_f)>\Omega(t_{f+1})$ is satisfied, that is, the time-step after which the angular velocity again begins to decrease. We then define the relative glitch size as $\Delta \Omega/\Omega = [\Omega(t_f)-\Omega(t_i)]/\Omega(t_i)$, and the waiting time $\Delta t$ as the time interval between $t_f$ for successive glitches.
\begin{figure}
	\includegraphics[width=\columnwidth]{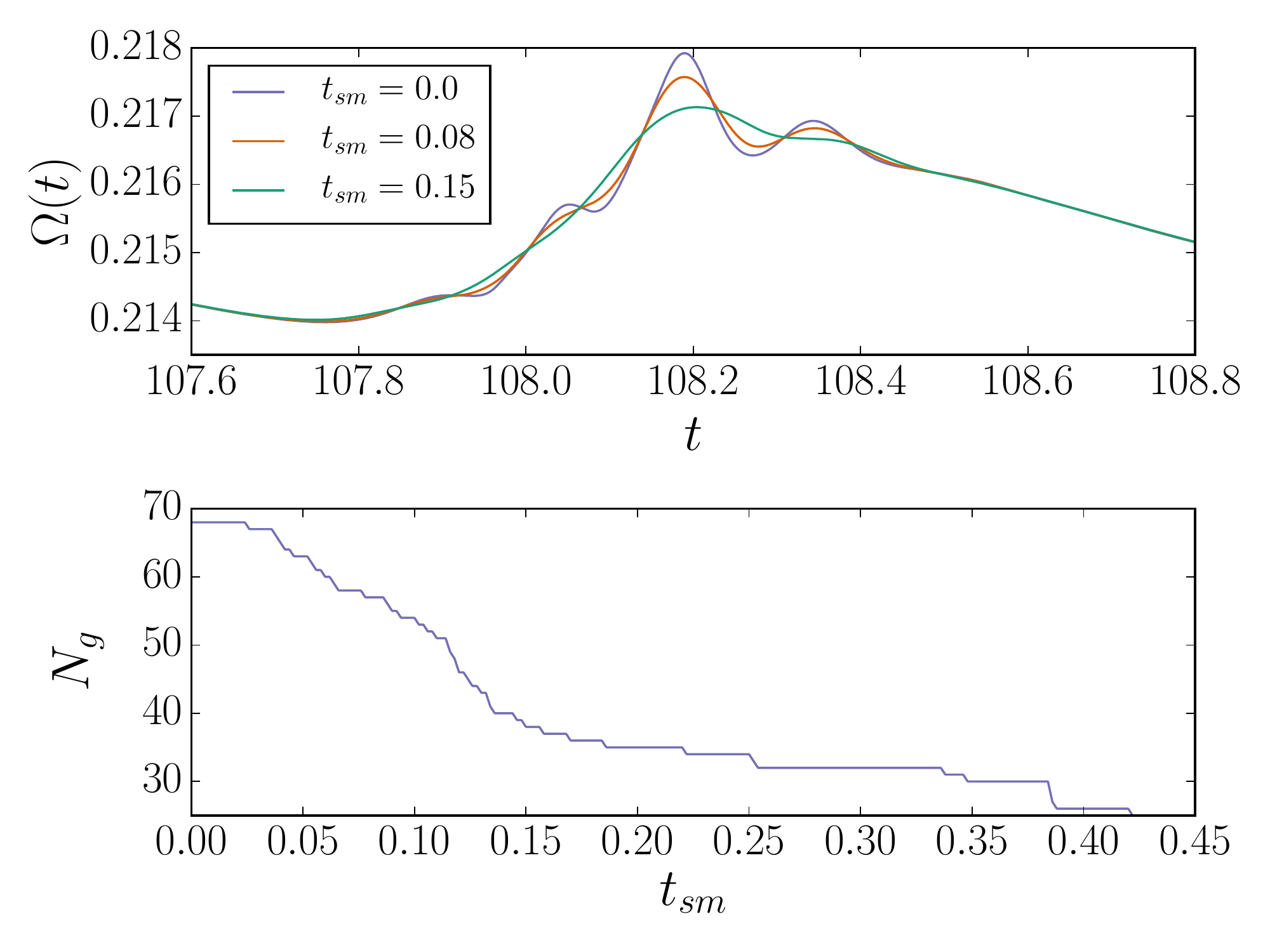}
    \caption{Features of the glitch-finding algorithm. Top panel: profile of a typical glitch for different smoothing time-scales $t_\mathrm{sm}$. For $t_\mathrm{sm} > 0.146$ only one glitch is detected in the plotted time interval, $107.6 \leq t \leq 108.8$, as required by eye. Bottom panel: number of glitches detected by the algorithm over the whole simulation ($0 \leq t \leq 190.8$) versus $t_\mathrm{sm}$. The top panel is a reproduction of Figure 5.6 in \protect\cite{Warszawski2011b}. Both plots are generated from the time series $\Omega(t)$. Parameters: $R=7$, $V_0/V_1 = 7/3$, $\xi=1.58$, $a_0=1$.}
    \label{fig:gl_algo}
\end{figure}
Figure \ref{fig:gl_algo} shows (top panel) the profile of a typical glitch, for different values of the smoothing time-scale $t_\mathrm{sm}$ and (bottom panel) the number of glitches detected by the algorithm over the whole simulation ($0 \leq t \leq 190.8$) versus $t_\mathrm{sm}$. As $t_\mathrm{sm}$ initially increases from $0$, a large number of small glitches are removed, while for $t_\mathrm{sm} \gtrsim 1.5$ the number of glitches decreases slowly with $t_\mathrm{sm}$. We conclude that the true number of glitches is approximately $40$, where $N_g$ flattens out, and take $t_\mathrm{sm}=0.15$. This coincides with the top panel, where for $t_\mathrm{sm} > 0.146$ the algorithm detects a single glitch in the time interval $107.6 \leq t \leq 108.8$, as required by eye and matching the time-scale over which the multiple peaks in the unsmoothed $\Omega(t)$ occur.
\subsection{Size and waiting time statistics}\label{sec:stats}
\begin{figure}
	\includegraphics[width=\columnwidth]{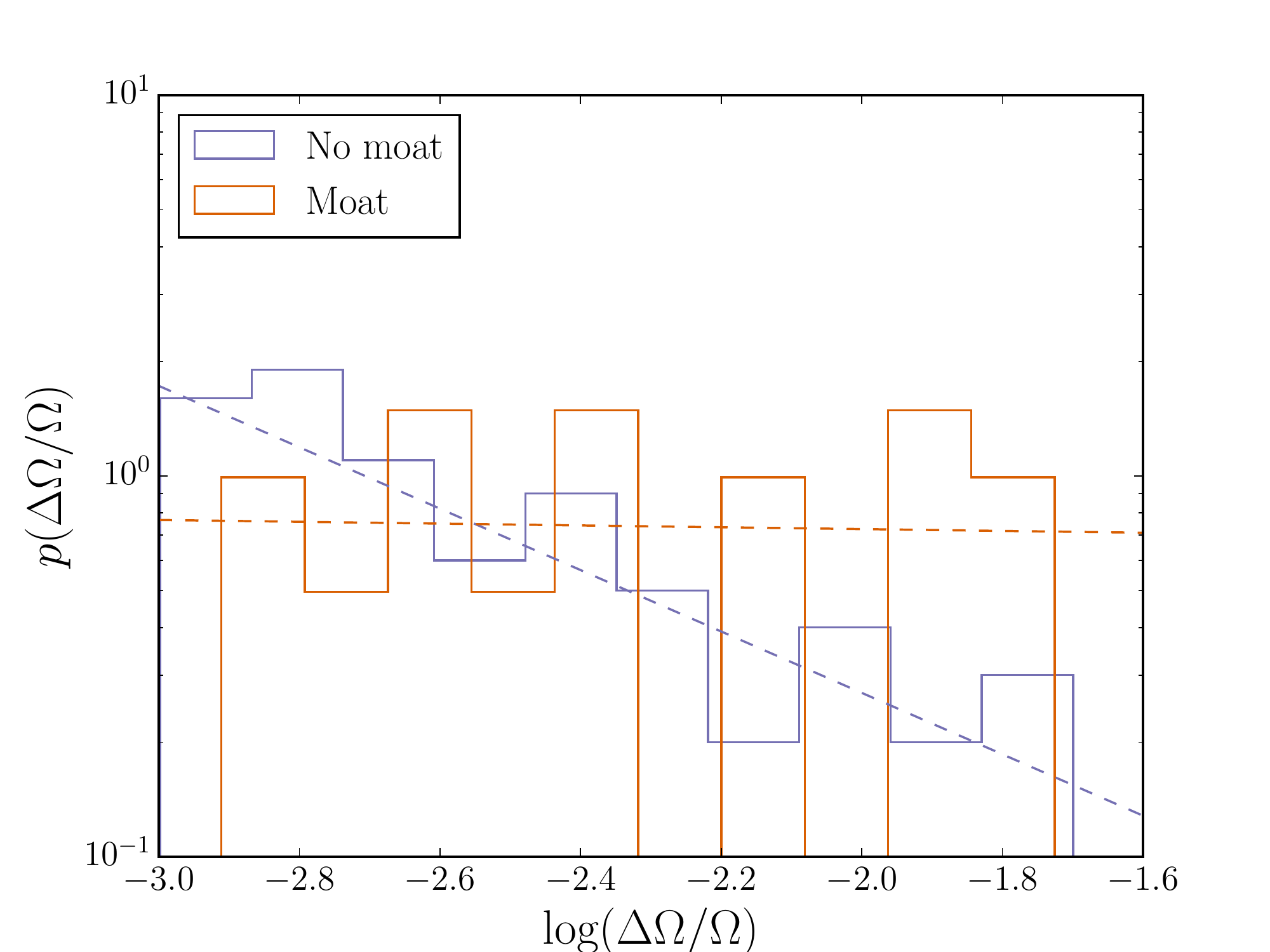}
    \caption{Probability density function (solid curves) of fractional glitch sizes $\Delta \Omega / \Omega$ with (orange) and without (purple) a moat. The dashed curves are power-law fits with $\alpha = -0.81$ (purple; no moat) and $\alpha = -0.02$ (orange; moat). We omit glitches with $\log(\Delta \Omega/\Omega)<-3$. Parameters: $V_0/V_1 = 10/3$, $\xi=1.58$, $a_0=1$, $t_\mathrm{sm} = 0.15$.}
    \label{fig:size_pdf}
\end{figure}

\begin{figure}
	\includegraphics[width=\columnwidth]{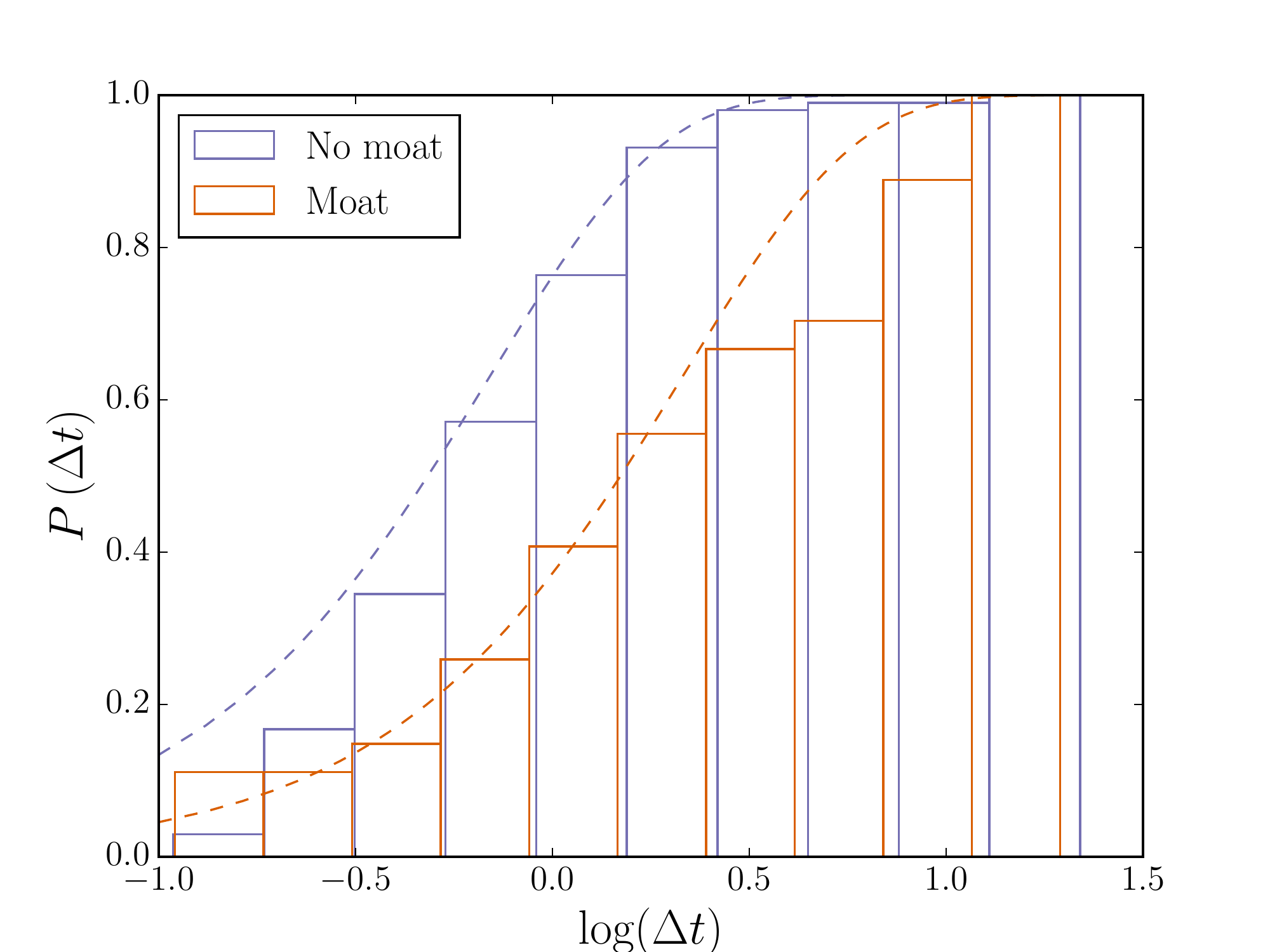}
    \caption{Cumulative probability distribution of waiting times $\Delta t$ with (orange) and without (purple) a moat. The dashed curves are exponential fits with $\lambda = 0.47$ (orange; moat) and $\lambda = 1.44$ (purple; no moat). A smaller $\lambda$ corresponds to less frequent glitches. Parameters: same as Figure \ref{fig:size_pdf}.}
    \label{fig:wait_pdf}
\end{figure}

The hypothesis that neutron star glitches are produced by avalanche dynamics implies that glitch sizes are distributed according to a power law probability density function $[p(\Delta \Omega/\Omega) \propto (\Delta \Omega/\Omega)^\alpha]$, and waiting times are distributed as an exponential $[p(\Delta t) = \lambda \exp(-\lambda \Delta t)]$ \citep{Jensen1998,Melatos2008}. This motivates the construction of probability density functions of these quantities from the simulations. Figure \ref{fig:size_pdf} shows the size probability density function $p(\Delta \Omega/\Omega)$ on $\log$-$\log$ axes with and without a moat (solid orange and purple curves respectively). The dashed curves are power-law fits with $\alpha = -0.02$ (moat) and $-0.81$ (no moat) over $\approx 1.4$ decades. We omit glitches with $\log{(\Delta\Omega/\Omega)}<-3$ as we are interested in the collective vortex dynamics and not small readjustments or `jiggling' involving few vortices. 
Figure \ref{fig:wait_pdf} shows the waiting time cumulative probability, $P(\Delta t) = \int_0^{\Delta t}d(\Delta t') p(\Delta t')$. An exponential fit to the data gives dimensionless mean glitch rates $\lambda = 0.47$ and $1.44$ with $(V_0 \neq 0)$ and without $(V_0 = 0)$ a moat respectively. The number of glitches detected by the algorithm is $N_g(V_0 \neq 0) = 28$ and $N_g(V_0=0)=204$. There are clear differences in the statistics with and without a moat. However, there are too few events to properly discriminate between a power law (say) and some other functional form. We do not claim that the data demonstrate a power-law size distribution; it is simply a convenient parametrization in keeping with previous work \citep{Warszawski2011a}.
\par
For $t_\mathrm{sm} = 0$ we have $N_g(V_0 \neq 0) = 48$ and $N_g(V_0=0)=592$, $\lambda(V_0\neq0) = 1.959$ and $\lambda(V_0=0) = 5.674$, and power-law indices $\alpha(V_0\neq0) = -0.302$ and $\alpha(V_0=0) = -1.098$. As discussed in Section \ref{sec:glitch-algorithm}, this case overestimates the true number of glitches by including multipeaked glitches and numerical jitter. For $t_\mathrm{sm} = 0.3$ we find $N_g(V_0 \neq 0) = 21$ and $N_g(V_0=0)=121$, $\lambda(V_0\neq0) = 0.282$ and $\lambda(V_0=0) = 0.944$, and $\alpha(V_0\neq0) = -0.031$ and $\alpha(V_0=0) = -0.476$. Although $N_g$, $\alpha$ and $\lambda$ depend on $t_\mathrm{sm}$, the overall shape of the probability density functions is similar. Moreover, the ordering of $N_g$, $\alpha$ and $\lambda$ is preserved between the moat and no moat cases for a wide range of $t_\mathrm{sm}$. Hence we can reasonably comment below on the qualitative effect of a moat on the statistics. We do this in the following paragraph. \par
The quantitative trends in the glitch statistics are as follows: compared to the no-moat system, a moat gives rise to (i) fewer glitches; (ii) a smaller mean glitch rate $\lambda$; and (iii) a smaller power-law index $\alpha$. That is, a moat gives rise to glitches which are larger but less frequent than those which occur in the absence of a moat. The values of $\alpha$ obtained in the two experiments differ by an order of magnitude, so that we can expect glitches to be significantly larger on average, if a moat is present.
The power-law index $\alpha=-0.81$ for an experiment without a moat but with a lattice of pinning sites is comparable to the range $-1.104 \leq \alpha \leq -0.994$, for different values of $t_\mathrm{sm}$, reported by \cite{Warszawski2011a}. The authors of that paper also reported, from exponential fits to $P(\Delta t)$, mean glitch rates of $0.011 \leq \lambda \leq  0.87$ (for different $t_\mathrm{sm}$) but find that these fits fail the Kolmogorov-Smirnov confidence test for the null hypothesis that the cumulative waiting time data are drawn from an exponential. \par
The results presented above are for $\gamma=0.1$ and $N_\text{ext}=-0.005$. For different $\gamma$ and $N_\text{ext}$, there is some variation in $\alpha$ and $\lambda$. For $\gamma$ in the range $0.05 \leq \gamma \leq 0.2$, we find $-0.96 \leq \alpha \leq 0.01$ and $0.61 \leq \lambda \leq 1.83$ with a moat. For $N_\mathrm{ext}=-0.01$, we find $\alpha=0.1$ and $0.33$ and $\lambda=0.42$ and $0.38$ with and without a moat respectively. Simulations with parameters well outside the above ranges were attempted but proved computationally intractable. In some runs few glitches are detected, so that an extra measure of caution should be taken when interpreting the results. Conducting a full study of the sensitivity of the size and waiting time statistics to $\gamma$ and $N_\mathrm{ext}$ lies beyond the scope of this paper.
\section{Conclusions}
In this paper, we study vortex motion in a rotating, decelerating BEC with a uniform grid of pinning sites plus an annular barrier (`moat') of deeper pinning, reminiscent of the set-up studied by \cite{Sedrakian1999}. The ultimate aim is to clarify the role of stratified pinning in a neutron star as input into future, idealized glitch models. However, one must exercise caution when interpreting the results astrophysically, because computational limitations force the simulations to be conducted under physical conditions far from those that exist in a neutron star, as discussed in Section \ref{parameters}. \par
We solve the time-dependent GPE and investigate the equilibrium vortex configuration, vortex dynamics and glitch statistics.
We find (Section \ref{sec:equilibrium}) that vortices pin preferentially in the moat, so that there is a vortex overdensity in the region $\lvert r-R \rvert < \xi$. The overdensity gives rise to large gradients in the azimuthal condensate velocity at $r \approx R$. The net outward vortex flux is unchanged by the moat, but the vortex flux out of the system as a whole is greater than the flux out of the region $\lvert r-R \rvert < \xi$ (Section \ref{flux}). In other words, the number of vortices pinned in the moat decreases with time but increases as a fraction of the total number of vortices in the system. The moat produces glitches which are fewer in number but on average larger than without a moat (Section \ref{dynamics}). \par
The study in this paper is motivated by the following specific astrophysical question: is it possible to detect the signature of stratified pinning in a neutron star in its long-term spin-down rate and glitch size and waiting time statistics? Needless to say, the results do not answer this question definitively, because the simulations are idealized in several important ways. For example, the dimensions of the simulation box are small, the number of vortices in a simulation is of order $100$ (compared to $10^{16}-10^{19}$ in a neutron star), and the ratio of pinning sites to vortices is of order $10$ [compared to $10^{10}$ in a neutron star; see \cite{Haskell2015b}]. Nevertheless, two results stand out as likely to be relevant astrophysically: the tendency for vortices to accumulate in a moat as the system evolves, and the reduction in number but increase in size of glitches when a moat is present. Larger GPE simulations containing more vortices and running for longer time intervals will be pursued in future work, although simulations approaching realistic neutron star conditions are beyond the reach of current computer technology.
\section*{Acknowledgments}
We thank Dr Tapio Simula for authorizing the use of his GPE solver in this paper. The code operates within the real-space product, finite-element, discrete-variable spatial representation and employs an explicit, fourth-order, split-operator technique to propagate the solution in time. The research was supported by funding from the Australian Research Council's Discovery Program. 
BH acknowledges support from Polish National Science Centre (NCN) grant SONATA BIS 2015/18/E/ST9/00577.
%


\bibliographystyle{mnras}
\bibliography{paper}



\appendix


\bsp	
\label{lastpage}
\end{document}